\documentclass[letterpaper, 10 pt, conference]{ieeeconf}  

\IEEEoverridecommandlockouts                              

\pdfoutput=1

\usepackage{graphicx}
\usepackage{color}
\usepackage{microtype}

\usepackage{ifthen}
\usepackage{amssymb}
\usepackage[normalem]{ulem} 
\usepackage[colorinlistoftodos, textwidth=2.5cm]{todonotes}

\usepackage{array}

\usepackage{listings}

\usepackage[bookmarks=true,pdfborder={0 0 1}]{hyperref}
\usepackage{url} 

\usepackage{balance}

\usepackage{fixltx2e}

\newboolean{showcomments}
\setboolean{showcomments}{true} 
\ifthenelse{\boolean{showcomments}}
  {
		\newcommand{\nbb}[2]{
		\fcolorbox{black}{yellow}{\bfseries\sffamily\scriptsize#1}
		{\sf$\blacktriangleright$\textcolor{blue}{\textit{#2}}$\blacktriangleleft$}
		}
		
		\newcommand{\remarks}[1]{\color{red}[#1]\color{black}}

		\newcommand{\del}[1]{\textcolor{red}{\sout{#1}}} 
  }
  {
		\newcommand{\nbb}[2]{}
		\newcommand{\remarks}[1]{}

		\newcommand{\del}[1]{} 
  }

\title{\LARGE \bf
Engineering the Hardware/Software Interface for Robotic Platforms --
A Comparison of Applied Model Checking with Prolog and Alloy
}

\author{Md Abdullah Al Mamun$^1$, Christian Berger$^1$ and J\"{o}rgen Hansson$^1$  
\thanks{$^{1}$Department of Computer Science and Engineering,
        Chalmers University of Technology and University of Gothenburg
        {\tt\small \{abdullah.mamun,christian.berger,
        jorgen.hansson\}@chalmers.se}}%
}

\begin{document}

\maketitle
\thispagestyle{empty}
\pagestyle{empty}

\begin{abstract}
Robotic platforms serve different use cases ranging from experiments
for prototyping assistive applications up to embedded systems for realizing
cyber-physical systems in various domains. We are using 1:10 scale
miniature vehicles as a robotic platform to conduct research in the domain
of self-driving cars and collaborative vehicle fleets. Thus, experiments
with different sensors like e.g.~ultra-sonic, infrared, and rotary encoders need
to be prepared and realized using our vehicle platform. For each setup, we need
to configure the hardware/software interface board to handle all sensors and
actors. Therefore, we need to find a specific configuration setting for each
pin of the interface board that can handle our current hardware setup
but which is also flexible enough to support further sensors or actors for
future use cases. In this paper, we show how to model the domain of the
configuration space for a hardware/software interface board to enable
model checking for solving the tasks of finding any, all, and the best possible
pin configuration. We present results from a formal experiment applying the
declarative languages Alloy and Prolog to guide the process of engineering the
hardware/software interface for robotic platforms on the example of a
configuration complexity up to ten pins resulting in a configuration space
greater than 14.5 million possibilities. Our results show that our domain model
in Alloy performs better compared to Prolog to find feasible solutions
for larger configurations with an average time of 0.58s. To find the best
solution, our model for Prolog performs better taking only 1.38s for the
largest desired configuration; however, this important use case is currently not
covered by the existing tools for the hardware used as an example in this
article.
\end{abstract}

\section{Introduction and Motivation}
\label{sec:Introduction}

Self-driving vehicles \cite{BBBC+08}, as one popular example for intelligent robotics,
highly depend on the usage of sensors of different kinds to automatically detect road
and lane-markings, detect stationary and moving vehicles, and obstacles on the
road to realize automated functionalities such as automatic driving and parking
or to realize collision prevention functions. New functionalities, based on market
demands for example, require the integration of new sensors and actors to
the increasingly intelligent vehicle. These sensors and actors are interfaced
by a hardware/software board, whose number of available physical connection
pins is however limited.

When selecting such an interface board for a robotic platform, we do not
necessarily limit our focus on a possible pin assignment for a set of
sensors/actors which is fulfilling our current needs. Additionally, we
consider also the possibility of extending the current hardware architecture
with additional sensors and actors using the same interface board for future
use cases. Furthermore, exchanging such an interface board might require
the modification of existing low-level code or requires the development of
new code for the embedded real-time OS to realize the data interchange with
the given set of sensors/actors.

\begin{figure*}[h!]
\begin{centering}
\includegraphics[width=\linewidth]{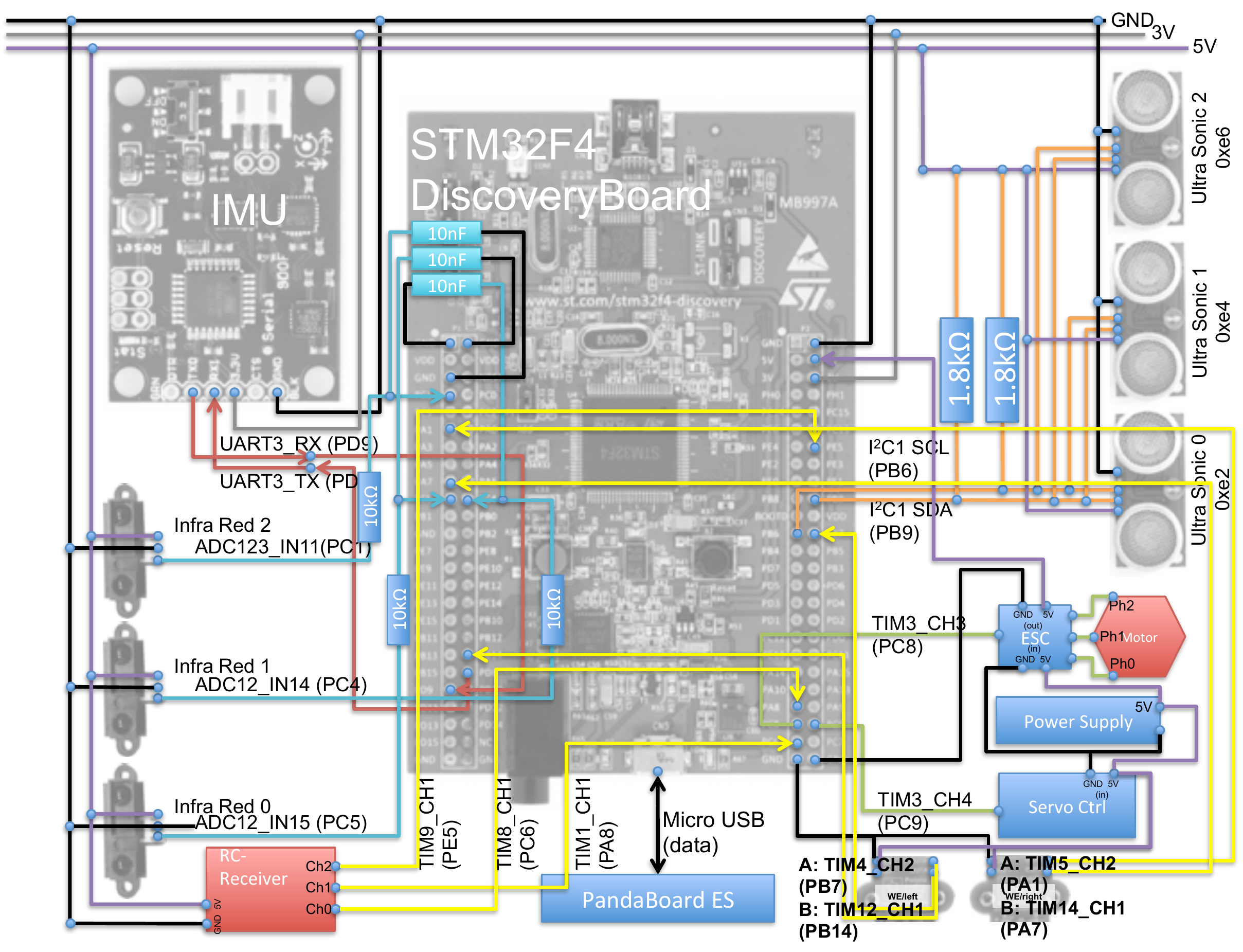}
\caption{Full setup of an STM32F4 Discovery Board with sensors and actors to realize the hardware/software interface for our self-driving miniature vehicle.}\label{fig:STM32F4BoardFullSetupReal}
\end{centering}
\end{figure*}

\begin{figure*}[ht]
\begin{centering}
\includegraphics[width=\linewidth]{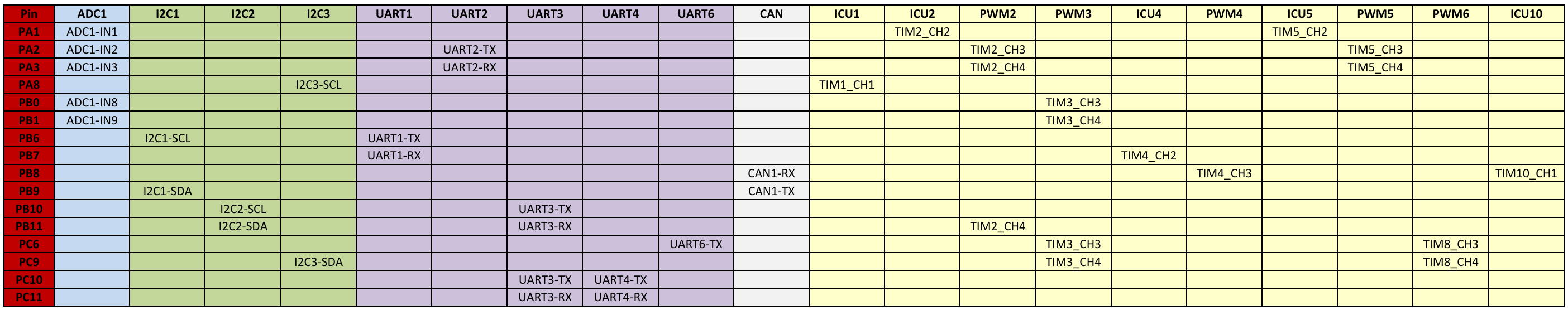}
\caption{Domain of possible pin assignment configurations for the STM32F4 Discovery Board: Analog input is marked with light blue, green highlights I$^2$C-bus usage, purple describes serial input/output, gray describes CAN bus connection, and light yellow ICU and PWM-timer-based input/output usage.}
\label{fig:pintable}
\end{centering}
\end{figure*}

As a running example in this paper, we are using the STM32F4 Discovery Board \cite{STM32}
as shown in Fig.~\ref{fig:STM32F4BoardFullSetupReal}. This figure depicts our
complete hardware/software interface setup for our self-driving miniature vehicle
consisting of different distance sensors, actors for steering and accelerating the
vehicle, an emergency stop over an RC-handset, as well as a connection to our
inertial measurement unit (IMU) to measure accelerations and angular velocities
for computing the vehicle's heading. The configuration space for that
interface board from which an optimal solution shall be chosen is shown in
Fig.~\ref{fig:pintable}.

The selection of an interface board of a certain type depends on different
factors like computation power and energy consumption. Furthermore, it must
support enough connection possibilities for the required sensors and actors.
However, matching a given set of sensors and actors to the available pins of
a considered hardware/software interface board is a non-trivial task because
some pins might have a multiple usage; thus, using one pin for one
connection use case would exclude the support of another connection use case.
To derive the best decision how to connect the set of sensors and actors, we
need to have a clear idea about all possible pin assignments up to a certain
length $l$, where $l$ describes the number of considered pins for one
configuration (e.g., a configuration length using ten pins could describe the
usage of 4 digital, 4 analog, and 2 serial pins).

From our experience, manually defining a feasible pin assignment for a desired
configuration requires roughly an hour, which includes checking the manual
and to evaluate, if future use cases for the HW/SW interface board can still
be realized. This process needs to be repeated, whenever the sensor layout is
modified, e.g. by adding further sensors or replacing sensors with different types
or replacing the existing interfacing board with a new one.
Thus, this manual work is time-consuming and error-prone.

Technical Debt is a recently promoted metaphor that uses concepts from 
financial debt to describe the trend of increasing software development 
costs over time. Manual tasks that can be repetitive over time and that 
have the possibility of being automated are a form of technical debt that 
accrues interest over time whenever a manual task is repeated \cite{TD_MLR_2013}. 
Thus automating the pin assignment configuration task would address 
challenges arising from technical debt.

In this paper, we address this configuration problem common for robotic
platforms by applying model checking to find a) at least one possible
pin assignment, b) all possible pin assignments, and consequently c) the
best possible pin assignment in terms of costs. In our case, costs are defined
as the number of multiple configurations per pin; e.g., let us assume one pin
from the hardware/software interface board can be used for analog
input, I$^2$C bus, and serial communication; its price would be 3. Reducing
the overall costs would result in a final pin assignment where pins with a low
multiple usage are preferred to allow for further use cases of the board in
the future.

We model the configuration problem as an instance of a domain-specific
language (DSL) for the configuration space of a hardware/software interface
board to serve the declarative languages Alloy \cite{Jackson2002} and
Prolog \cite{CM03}. Based on this model, we show how to realize
the aforementioned three use cases in these languages while measuring
the computation time to compare the model checking realized by these tools.

The rest of the paper is structured as follows: In Sec.~\ref{sec:sensorlayoutverification},
the combinatorial optimization problem for engineering the
hardware/software interface board is introduced, formalized, and
constraints thereof are derived. Furthermore, the complexity of the
configuration space is analyzed before the formal experiment of
applying model checking with Alloy and Prolog and its results are
described, analyzed, and discussed. The article closes with a discussion of
related work and a conclusion.

\section{Engineering The Robotic Hardware/Software Interface--A Combinatorial Optimization Problem}
\label{sec:sensorlayoutverification}

Fig.~\ref{fig:STM32F4BoardFullSetupReal} shows the connection setup of the
hardware/software interface board that we are using on our 1:10 scale self-driving
miniature vehicle platform. In the given configuration, the board handles 14
different input sources and two output sinks:

\begin{itemize}
\item three Sharp GP2D120 infrared sensors which are generating a distance-dependent voltage level,
\item an IMU Razor 9DoF board connected via a serial connection that provides acceleration and angular velocity data in all three dimensions as well as housing a magnetometer to provide information about the vehicle's heading,
\item a three-channel receiver for the remote controller handset to stop and control the miniature vehicle in emergency cases connected as analog source to the input capturing unit (ICU),
\item three (and up to 16) ultra-sonic devices attached via the I$^2$C digital bus,
\item and a steering and acceleration motor connected via pulse-width-modulation (PWM) pins to access the actors of the robotic platform.
\end{itemize}

To handle all aforementioned sensors and actors using ChibiOS \cite{chibios}
as our hardware abstraction layer (HAL) and real-time operating system, we
need to engineer both the hardware connection mapping as well as the software
configuration setup fulfilling the following constraints:

\begin{itemize}
\item Attaching the hardware data sources to those pins that are able to handle the required input source at hardware-level (e.g.~the STM32F4 chip in our case),
\item connecting the hardware data sinks to those pins that are able to handle the required output sources at hardware-level,
\item configuring the CPU to handle the hardware data sources and sinks in the case of multiple usage per pin,
\item and considering the appropriate software support in the low-level layer of the hardware abstraction layer (e.g.~in our case considering that ICUs can only be handled if attached to a pin supporting timers on channel 1 or 2).
\end{itemize}

The aforementioned constraints need to be considered during the engineering
process. In this section, we describe the general idea behind our modeling approach
for these domain-specific constraints, considerations about the complexity in the
model processing stage, as well as how instances of the DSL are transformed to
enable model checking serving the following use cases during the engineering
process for the hardware/software interface board:

\begin{enumerate}
\item Find a feasible and valid pin configuration fulfilling a requested configuration,
\item enumerate all possible pin configurations for a given configuration,
\item and in combination with the former use case, find the best possible pin configuration in terms of costs for pin usage.
\end{enumerate}

\subsection{The Domain of Pin Assignment Configurations}
\label{subsec-domain}

\begin{figure}[ht!]
\begin{centering}
\includegraphics[width=\linewidth]{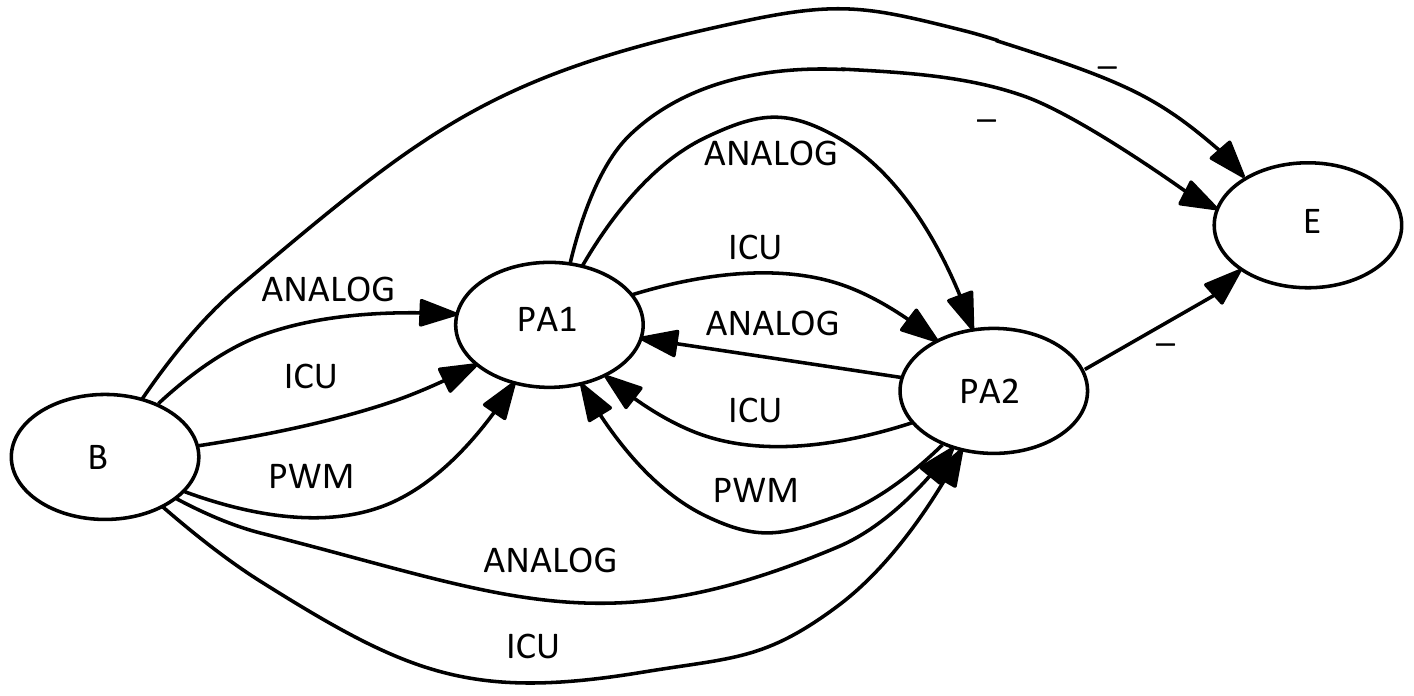}
\caption{Visualization of the graph $G = \{N,E,A\}$ for the domain of possible pin assignment configurations of a fictitious hardware/software interface board with two pins having multiple usage respectively. A concrete configuration is represented by a path $P$ from $n_B$ to $n_E$ with $|P| < |N|$.}\label{fig:configurationSpace}
\end{centering}
\end{figure}

In Fig.~\ref{fig:configurationSpace}, a visualization for the domain of possible
pin assignment configurations is depicted. The basic model can be represented
by a graph $G$ consisting of nodes $N$ representing all pins of a hardware/software
interface board, a set $E$ describing directed edges connecting the nodes, and a
set $A$ of edge annotations representing concrete pin configurations. One concrete
pin assignment configuration is then represented by a path $P$ from $n_B$ to $n_E$.

Furthermore, the following constraints must hold to restrict the set of possible
paths through $G$ to consider only those representing valid configurations:

\begin{itemize}
\item The graph must not contain self-reflexive edges at the nodes because one pin can only be used once for a pin configuration usage.
\item The path $P$ of a concrete pin assignment configuration must begin at $n_B$ and must end in $n_E$.
\item The length of $P$ must be less than the size of set $N$.
\end{itemize}

\begin{figure*}[h!]
\begin{centering}
\includegraphics[width=0.8\linewidth]{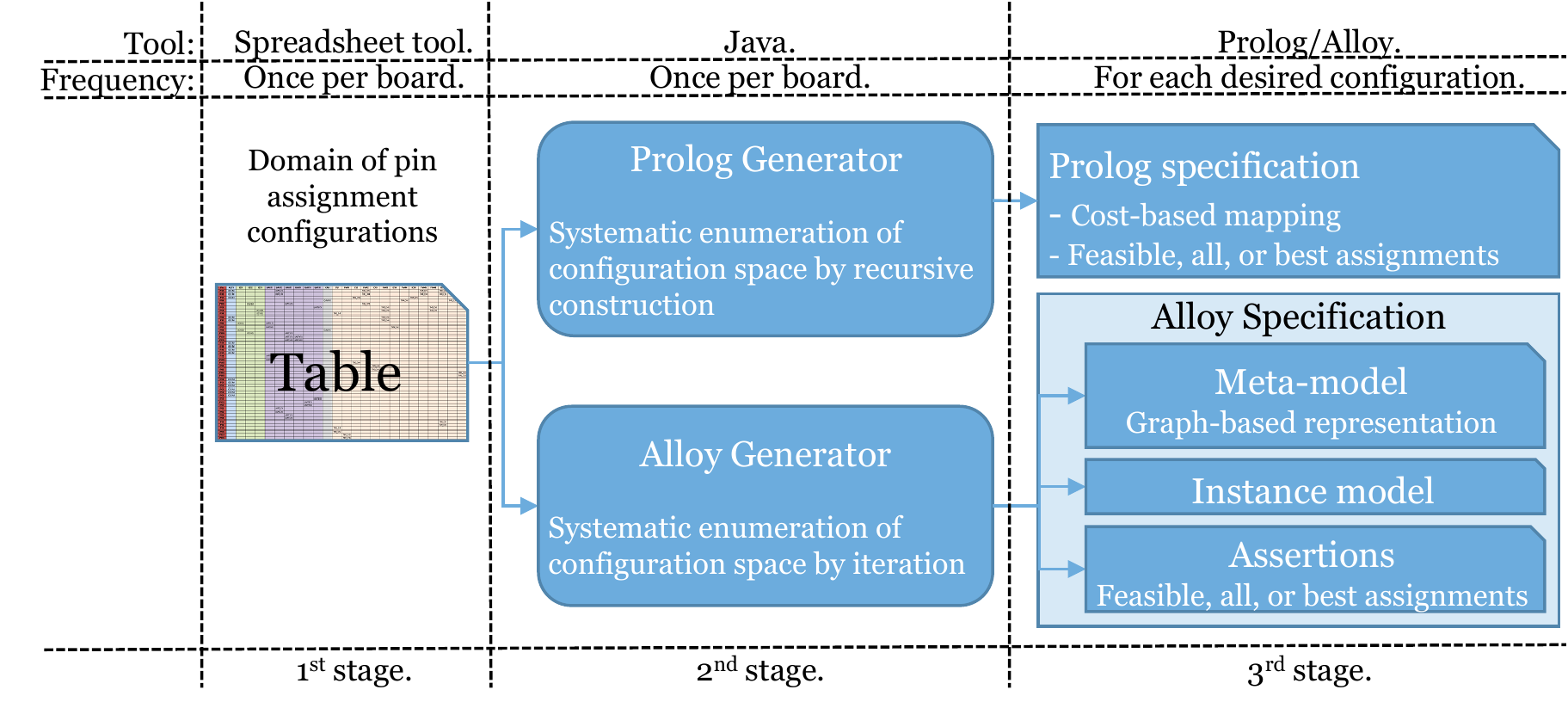}
\caption{Overview of the workflow with involved tools for transforming an instance of the domain model for possible pin assignment configurations into representations, which can be used for model checking and to compute a feasible, all possible, or the best pin assignment for a hardware/software interface board. The first two stages need to be maintained once per hardware/software interface board, while the last stage needs to be carried out for each desired configuration.}
\label{fig:workflow}
\end{centering}
\end{figure*}

This domain-specific model can also be represented as a table as shown in
Fig.~\ref{fig:pintable}, which can be maintained with any
spreadsheet tool for example. Thus, only all possible configuration settings need to be
defined per pin because all aforementioned constraints must be
considered only during the concrete assignment process, which in turn can
be fully automated with model checking. An overview of the model checking
workflow is shown in Fig.~\ref{fig:workflow}.

The concrete realizations for both paths in the workflow are described in
Sec.~\ref{subsec:prolog} for Prolog and in Sec.~\ref{subsec:alloy} for Alloy.

\subsection{Complexity Considerations}

The combinatorial complexity of finding a solution for the pin assignment
problem for a given configuration with a length $l$ is determined by the
following three dimensions: Set $N$ of available pins, set $M$ of different
configurations per pin, and the maximum length $L$ up to which
the assignments shall be solved.

For $|M| = 1$, the combinatorial problem is reduced to determining
how many possibilities $C_{|M|=1}$ are available to pick $k$
objects from $N$ as calculated by the binomial coefficient shown
in Eq.~\ref{ConfigurationsM=1}. Hereby, $k$ describes the length of
a considered configuration.

\begin{equation}
C_{|M|=1}^{|N|} = \sum_{k=1}^{L} {{|N|}\choose {k}} = \sum_{k=1}^{L} \frac{|N|!}{(|N|-k)! k!}.
\label{ConfigurationsM=1}
\end{equation}

However, the configuration space grows once the limitation for set $M$
is relaxed as outlined in the following example:

\begin{eqnarray}
C_{|M|=1}^{|N|=4} & = & 4 + 6 + 4 + 1 = 15.  \nonumber \\
C_{|M|=2}^{|N|=4} & = & 4 * 2 + 6 * 3 + 4 * 5 + 1 * 6 = 47. \nonumber \\
C_{|M|=3}^{|N|=4} & = & 4 * 3 + 6 * 6 + 4 * 10 + 1 * 15 = 103. \nonumber \\
& & \dots \nonumber
\end{eqnarray}

Analyzing the factors, which are multiplied with the binomial coefficient
summands, it can be seen that they are constructed by the rule depicted
by Eq.~\ref{Factors}.

\begin{equation}
K(n, m) = \left\{ \begin{array}{ll}
 1 + \sum_{p=1}^{n} * K(p, m-1) &\mbox{if $m>1$,} \\
  1 &\mbox{otherwise.}
       \end{array} \right.
\label{Factors}
\end{equation}

Using Eq.~\ref{Factors}, Eq.~\ref{ConfigurationsM=1} can be adapted for the
generic case as shown in Eq.~\ref{ConfigurationsM}.

\begin{equation}
C_{|M|}^{|N|} = \sum_{k=1}^{L} {{|N|}\choose {k}} * K(k, |M|).
\label{ConfigurationsM}
\end{equation}

With Eq.~\ref{ConfigurationsM}, a hardware/software interface board
consisting of 6 pins each providing 4 different configuration possibilities would
result in 1,519 different assignment options.

\section{Evaluating Applied Model Checking for Pin Assignment Configurations}

In the previous section, we have outlined the domain of possible pin assignment
configurations alongside with complexity considerations. Now, we investigate
the following research questions related to the challenges during the engineering
process of the hardware/software interface for robotic platforms:

\begin{description}
\item[\textit{RQ-1:}] How can Prolog be used to apply model checking on instances of the domain of possible pin assignment configurations to determine a feasible, all possible, and the best configuration assignment?
\item[\textit{RQ-2:}] How can Alloy be used to apply model checking on instances of the domain of possible pin assignment configurations to determine a feasible, all possible, and the best configuration assignment?
\item[\textit{RQ-3:}] Which approach performs better compared to the other for the particular use cases?
\end{description}

Since we have full control over the involved parameters for the model checkers,
we carried out a formal experiment according to \cite{Pfl94} to answer the research questions.

\subsection{Designing the Formal Experiment}

To compare the possibilities and performance of Prolog and Alloy, both tools
were used to solve the following problems:

\begin{enumerate}
\item Basis for the formal experiment was the concrete instance of possible pin assignment configurations for the 46 pins of our hardware/software interface board STM32F4 Discovery board.
\item From this instance, 30 trivial assignments with costs of 1 were removed because any identified assignment for the pins with multiple usage can be simply extended by pins with costs 1 without modifying the assignment for the other pins.
\item For the remaining 16 pins as shown in Fig.~\ref{fig:pintable} which can be used with multiple configurations up to costs of 4, a pin assignment for a given configuration of varying lengths ranging from 0 up to 10 is needed to be solved for the use cases \textit{one feasible}, \textit{all possible}, and \textit{the best} pin assignment.
\item The given configuration, for which pin assignments are needed to be determined, consisted of \texttt{\{analog, analog, analog, icu, analog, analog, serial-tx, serial-rx, can-tx, i2c-sda\}}. This list was shortened from the end to provide shorter configurations as input.
\item To verify that both model checking approaches identify also impossible configurations, a given configuration containing too many elements from a given type of set $A$ was constructed.
\item For every use case and for every configuration length, the required computation time was determined.
\end{enumerate}

To answer RQ-1 and RQ-2, respectively, we decided to use action design research
\cite{MOS+11} as the method to identify and analyze a domain problem for designing
and realizing an IT artifact to address the problem.

To answer RQ-3, we decided to measure the \textit{required computation time}
for each approach because in our opinion, it is the apparent influencing factor for
the last stage in our workflow, where researchers and developers have to cope
with during the development and usage of a robotic platform.

According to Eq.~\ref{ConfigurationsM}, the total configuration space for the
running example with $|N| = 16$ pins and $|M| = 20$ configuration possibilities
would contain 1,099,126,862,792 elements. However, due to the reduced number
of multiple usages per pin in our concrete example of the STM32F4 Discovery Board,
this space is reduced to 14,689,111 possibilities.

In the following, the formal experiment with Prolog and Alloy is described respectively.

\subsection{Verification Approach Using Prolog}
\label{subsec:prolog}

\textbf{Target Model Design}

This approach uses the logic programming language Prolog \cite{CM03} to
verify a given input configuration for the hardware/software interface. Prolog
is a declarative language based on Horn clauses. Our target model which we
derive from the tabular input specification consists of facts and an inference
part. A fact in our model describes hereby a possible configuration as a mapping
from the given configuration assignment to a pair consisting of a list of
specific pins realizing this configuration and the associated costs like the
following: \texttt{config([analog,analog],[[pa1,pa2],7]).} This fact describes
that pins \texttt{pa1} and \texttt{pa2} can be used to serve two analog inputs
with the associated costs of 7.

\begin{figure}[h!]
\begin{verbatim}
getConfig(RequiredConfiguration, Pair) :-
    msort(RequiredConfiguration, S),
    config(S, Pair).

allConfigs(RequiredConfiguration, Set) :-
    setof([Pins,Costs],
        getConfig(RequiredConfiguration,
        [Pins,Costs]), Set).

cheapestConfig(R, Pins, Costs) :-
    setof([Pins,Costs],
     getConfig(R, [Pins,Costs]), Set),
    Set = [_|_],
    minimal(Set, [Pins,Costs]).
\end{verbatim}
    \caption{Excerpt from the inference rules from our Prolog model.}
    \label{fig:inference}
\end{figure}

An excerpt of the inference rules is shown in Fig.~\ref{fig:inference}
providing the interface to the the target model. Hereby, we have the
methods to get one feasible (\texttt{getConfig/2}), all possible
(\texttt{allConfigs/3}), and the best pin assignment
(\texttt{cheapestConfig/3}). Due to optimization reasons, the facts
and inference rules are instantiated for the particular lengths of
given configurations.

\textbf{Model Transformation \& Constraint Mapping}

To transform our input specification from the domain of possible pin
assignment configurations to the target model in Prolog, we have realized
the model transformation in Java. Hereby, the algorithm recursively traverses
the tabular representation to create a hashmap with an ordered list of a
configuration assignment as a key and a list of possible pins realizing
this assignment as the associated value to the key. Due to the internal
order of the used keys, the set of identified possible configurations was
reduced. However, this design decision would require that the user would
need to specify an ordered configuration request to find a suitable match
from the facts; to relax this constraint, Prolog's function \texttt{msort/2}
was incorporated to sort any request before it is actually evaluated while
preserving duplicates. Furthermore, during the table traversal, the constraints
as listed in Sec.~\ref{subsec-domain} are obeyed to avoid self-reflexive pin
assignments or resulting configurations using more pins than available.

The resulting hashmap is then iterated to create the single facts for Prolog
by resolving the keys to the list of associated pins realizing this configuration.
During this step, the specific costs for a concrete pin assignment are also
determined. Generating the target model and applying the constraints during
the traversal process took approximately 2,102.4s. These processing steps need to be
done only once per hardware/software interface board since the actual model
checking is realized in Prolog afterwards.

\textbf{Results}

In the following, the results from our experiment applying model checking
with Prolog are presented. In Table \ref{tbl:PrologFeasibleAssignment}, the
costs for one feasible pin assignment alongside with the Prolog computation
time for different configuration lengths from 1 to 10 are shown. This table
also shows the computation times for impossible configurations.

\begin{table}[htb]
\begin{center}
\begin{tabular}{ c || >{\centering\arraybackslash}m{1.5cm} | >{\centering\arraybackslash}m{2.1cm} || >{\centering\arraybackslash}m{2.1cm}}
Length & Costs for feasible assignment & Computation time for feasible configuration & Computation time for impossible configuration \\
\hline
\hline
1 & 3 & 0s & 0s \\
\hline
2 & 7 & 0s & 0s \\
\hline
3 & 11 & 0s & 0s \\
\hline
4 & 13 & 0s & 0.01s \\
\hline
5 & 15 & 0.03s & 0.02s \\
\hline
6 & 17 & 0.11s & 0.10s \\
\hline
7 & 19 & 0.29s & 0.30s \\
\hline
8 & 21 & 0.78s & 0.64s \\
\hline
9 & 23 & 1.06s & 1.06s \\
\hline
10 & 26 & 2.47s & 1.36s \\
\hline
\hline
 &  & $\varnothing = $0.474s  & $\varnothing = $0.349s  \\
 &  & $\pm$ 0.79s             & $\pm$ 0.50s             \\
\hline
\end{tabular}
        \caption{Prolog results to check both possible and impossible pin configurations
        for different configuration lengths.}
        \label{tbl:PrologFeasibleAssignment}
\end{center}
\end{table}

Table \ref{tbl:PrologAllAndBestAssignment} shows the results to
find all possible pin assignment configurations and among them,
also the best assignment in terms of costs for different configuration
lengths from 1 to 10. If the identified pin assignment solution is
cheaper compared to the previous table, the costs are highlighted.

\begin{table}[htb]
\begin{center}
\begin{tabular}{ c || c | c | c }
Length & Number of   & Costs for best & Prolog computation \\
       & all possible & assignment     & time (all/best)\\
       & assignments  &                 &               \\
\hline
\hline
1 & 5 & \textbf{2} & 0s/0s \\
\hline
2 & 10 & \textbf{4} & 0s/0s \\
\hline
3 & 10 & \textbf{7} & 0s/0s \\
\hline
4 & 24 & \textbf{9} & 0.01s/0.01s \\
\hline
5 & 11 & \textbf{13} & 0.06s/0.03s \\
\hline
6 & 2 & 17 & 0.22s/0.11s \\
\hline
7 & 8 & 19 & 0.61s/0.30s \\
\hline
8 & 20 & 21 & 1.40s/0.64s \\
\hline
9 & 20 & 23 & 2.42s/1.08s \\
\hline
10 & 32 & 26 & 4.06s/1.38s \\
\hline
\hline
 &  & & $\varnothing_{all}$ = 0.878s $\pm$ 1.375s \\
 &  & & $\varnothing_{best}$ = 0.355s $\pm$ 0.51s \\
\hline
\end{tabular}
        \caption{Results to check for all possible as well as the best pin assignment for different configuration lengths. If a better pin assignment in terms of costs was found compared to Table \ref{tbl:PrologFeasibleAssignment}, the entry is highlighted.}
        \label{tbl:PrologAllAndBestAssignment}
\end{center}
\end{table}

\subsection{Verification Approach Using Alloy}
\label{subsec:alloy}

\textbf{Target Model Design}

This approach uses Alloy \cite{Jackson2002} to verify the input configuration
space of the hardware/software interface. Alloy is a declarative language
influenced by the Z specification language. Alloy expressions are based on first
order logic and models in Alloy are amenable to fully automatic semantic
analysis. However, Alloy does not perform fully exhaustive analysis of the
models but rather makes reductions to gain performance.

We have used assertions in Alloy to verify whether a certain configuration
is viable in the hardware/software interface board. Checking assertions results
either true or false reflecting the unsatisfiability of the given predicate. If a
predicate is not satisfiable, the Alloy analyzer reports counterexamples showing
how the predicate is invalid.

To use Alloy for model checking, we transform the tabular input specification
into an equivalent representation as described by a meta-model consisting of
classes \emph{Pin, ConnType, ConnDetail}, and \emph{Cost}
and references \emph{conntype, conn\_detail}, and \emph{cost} originating
from \emph{Pin} with mapping cardinalities 0 - 1..*, 0 - 1..*
and 1 - 1 respectively to the respective classes. Hereby, \emph{Cost} is a
derived construct originally not available in the input specification.

\textbf{Model Transformation \& Constraint Mapping}

A given instance model conforming to the meta-model alongside
with the domain constraints as listed in Sec.~\ref{subsec-domain} is transformed
to an Alloy specification. This instance model defines
Alloy signatures for all connection types, connection details
and pins available in the input specification. Two signatures from
the specification are shown in Fig.~\ref{fig:alloy-spec}.

\begin{figure}[h!]
\begin{verbatim}
one sig PA1 extends Pin {} {
  conntype = ANALOG + ICU + ICU
  conn_detail = ADC1_IN1 + TIM2_CH2 + TIM5_CH2
  cost = 3}

one sig PA2 extends Pin {} {
  conntype = ANALOG + SERIAL_TX + ICU + ICU
  conn_detail = ADC1_IN2 + UART2_TX +
  TIM2_CH3 + TIM5_CH3
  cost = 4}
\end{verbatim}
    \caption{Alloy instance specification for two pins.}
    \label{fig:alloy-spec}
\end{figure}

Checking Alloy assertions can find a feasible pin assignment for a given
configuration. Assertions in Alloy may report counterexamples showing violations
of the assertions with respect to the specification facts. Since, we want to
find out a possible pin configuration, we generate assertions in Alloy assuming
that the inverse statement of that request would be true. Then, we let Alloy
find a counterexample, which in turn represents a possible realization of the
desired configuration. An example for such a negated statement is depicted
in Fig.~\ref{fig:genalloyassertion}.

\begin{figure}[h!]
\begin{verbatim}
assert ANALOG_ANALOG {
	all disj p1, p2:Pin |
	not (
	   ANALOG in p1.conntype &&
	   ANALOG in p2.conntype
	)}

check ANALOG_ANALOG
\end{verbatim}
    \caption{Generated negated assertion for the desired configuration ``\emph{ANALOG, ANALOG}''.}
    \label{fig:genalloyassertion}
\end{figure}

If Alloy succeeds to find a counterexample, the variables \texttt{p1} and
\texttt{p2} contain a feasible assignment to the pins of the hardware/software
interface board. We have dealt with two ways of generating Alloy assertions.
First, assertions for finding a feasible pin assignment for a desired configuration.
This follows a trivial solution of reading and transforming the input string
into Alloy expressions similar to the Fig.~\ref{fig:genalloyassertion}.

Second, assertions for finding the best possible solution. Alloy does not
support higher order quantification to write predicates or assertions, which
can automatically compute the cheapest possible pin assignment for a certain
configuration of a specific length. Thus, we have generated a series of assertions
where each of the assertions explores the possibility of a pin assignment for a
specific total cost level. If we consider a domain of possible pin assignments with
a minimum pin cost  \emph{PC\textsubscript{min}} and maximum pin cost
\emph{PC\textsubscript{max}}, then for a desired configuration of length
$l$, we have generated in total $l\times$\emph{PC\textsubscript{max}}~-~$l\times$\emph{PC\textsubscript{min}}
assertions. We have written a Java program to iteratively call these assertions
within a cost-range starting from the cheapest possible cost for the desired
configuration (i.e., $l\times$\emph{PC\textsubscript{min}})
to the maximum possible cost (i.e., $l\times$\emph{PC\textsubscript{max}})
and we stop the iteration as soon as we have found a solution.

Assertions for computing the best possible solution differ from the assertion
in Fig.~\ref{fig:genalloyassertion}. To enable this use case, we added the
expression ``\emph{p1.cost.add[p2.cost]$<$=X}'' where \emph{X} is taking
a total cost value within the  range mentioned above inside the \emph{not()} expression
of the assertion and by specifying integer bit-width in the corresponding
\emph{check} statement.

To generate the Alloy specification from the domain model, our Java
program took approximately 0.3s. This step needs to be done only
once per hardware/software interface board.

\textbf{Results}

The results of possible and impossible desired pin configuration are presented in
Table~\ref{tbl:AlloyFeasible-impossibleAssignment} showing costs and computation
time for possible and impossible configurations.
Table~\ref{tbl:AlloyAllAndBestAssignment}
shows results for all and best pin assignments for possible desired configurations.
A cost in these tables is a sum of all the costs of the pins associated with the
solution of the desired configuration. The sum of the costs is not automatically
processed by Alloy. However, it would be possible to post-process the output
data to automatically compute the costs.

\begin{table}[htb]
\begin{center}
\begin{tabular}{ c || >{\centering\arraybackslash}m{1.22cm} | >{\centering\arraybackslash}m{2.3cm} || >{\centering\arraybackslash}m{2.3cm}}
Length & Costs for the first feasible assignment & Computation time for feasible configuration & Computation time for impossible configuration \\
\hline
\hline
1 & 3 & 0.53s & - \\
\hline
2 & 7 & 0.52s & 0.52s \\
\hline
3 & 11 & 0.56s & 0.53s \\
\hline
4 & 13 & 0.54s & 0.53s \\
\hline
5 & 15 & 0.56s & 0.53s \\
\hline
6 & 17 & 0.57s & 0.64s \\
\hline
7 & 19 & 0.62s & 0.62s \\
\hline
8 & 22 & 0.63s & 0.56s \\
\hline
9 & 23 & 0.65s & 0.67s \\
\hline
10 & 26 & 0.67s & 0.68s \\
\hline
\hline
 &  & $\varnothing = $0.58s $\pm$ 0.05s& $\varnothing = $0.59s $\pm$ 0.06s\\
\hline
\end{tabular}
        \caption{Alloy results to check both possible and impossible configurations
        for different lengths.}
        \label{tbl:AlloyFeasible-impossibleAssignment}
\end{center}
\end{table}

\begin{table}[htb]
\begin{center}
\begin{tabular}{ >{\centering\arraybackslash}m{0.7cm} || >{\centering\arraybackslash}m{1.8cm} | >{\centering\arraybackslash}m{1.5cm} | >{\centering\arraybackslash}m{2.7cm} }
Length & Number of all possible assignments & Costs for best assignment & Alloy computation time (all/best) \\
\hline
\hline
1 & 5 & \textbf{2}  & 0.07s/0.53s \\
\hline
2 & 20 & \textbf{4}  & 0.24s/0.63s \\
\hline
3 & 60 & \textbf{7} & 0.59s/0.67s \\
\hline
4 & 480 & \textbf{9} & 1.57s/1.63s \\
\hline
5 & 840 & \textbf{13} & 2.27s/1.20s \\
\hline
6 & 720 & 17          & 2.16s/1.17s \\
\hline
7 & 2760 & 19          & 4.68s/1.09s \\
\hline
8 & 7320 & \textbf{21} & 10.43s/3.25s \\
\hline
9 & 7320 & 23          & 9.27s/2.88s \\
\hline
10 & 9960 & 26         & 14.12s/3.38s \\
\hline
\hline
 &  &  & $\varnothing_{all}$ = 4.58s $\pm$ 5.02s \\
 &  &  & $\varnothing_{best}$ = 1.64s $\pm$ 1.11s \\
\hline
\end{tabular}
        \caption{Results to check for all possible as well as the best pin assignment for different configuration lengths. If a better pin assignment in terms of costs was found compared to Table \ref{tbl:AlloyFeasible-impossibleAssignment}, the entry is highlighted.}
        \label{tbl:AlloyAllAndBestAssignment}
\end{center}
\end{table}

\subsection{Analysis and Discussion}

The results show that with Alloy, the growth of the computation time
with respect to the increasing lengths of the desired configurations
is moderate both for finding a feasible solution and for computing
an impossible configuration.
On the other hand, Prolog performs better on finding all and best
pin assignments for a desired possible configuration.
However, in the given scope of this experiment, both Prolog and Alloy
not only are able to find solutions for all of the outlined use cases but
also reporting the same solution with the same costs for finding the
best pin  assignment for a possible desired configuration.

The reason behind the surprisingly higher number of solutions reported
by Alloy for all possible solutions is that the generated Alloy assertions
report solutions that are not unique with respect to the pins.

From a practical point of view, finding a best pin assignment for a desired
possible configuration is more valuable than feasible and all pin
assignments. To find the best pin assignment, both Prolog and Alloy computation
times increase by the length of the configuration. In this case, the growth of
Prolog is smaller than the one from Alloy which ranks Prolog more scalable
with the size of the configuration length compared to Alloy under the terms
of settings for our experiment.

Furthermore, the Prolog solution provides a better user interaction in terms of
taking input configuration requests and producing corresponding output.
Moreover, the Prolog solution calculates the costs automatically, which is not
inherently supported by Alloy but possible to achieve with work-around
solutions.

Concerning the generation of the target model specification in the second
stage of our workflow, Alloy takes considerably less time and space compared
to Prolog. The size of the Alloy specification is less than 100KB compared to
1.7GB for Prolog. Loading the Alloy specification happens nearly instantly,
while loading and compiling the target model specification for Prolog to
start the model-checking process took 346.99s.

\subsection{Threats to Validity}

We discuss threats to validity to the results of our experiment according to the
definition reported by Runeson and H\"ost \cite{RH08}:

\begin{itemize}

\item \textit{Construct validity.} With respect to RQ-1, the outlined approach
with Prolog showed a possibility to apply model checking to verify a given
configuration and to find a feasible, all possible, and the best pin assignment
for a problem size, where researchers and engineers working with robotic
platforms are faced with.

As RQ-2 mentions, the solution with Alloy is also able to find a feasible, all,
the and best pin assignment for a specific pin configuration. A check statement
in Alloy does not guarantee that the associated assertion is invalid, if it does not
report a counterexample unless the scope of the check is proper. We have taken
necessary measures so that the scope always covers all possible solutions.
For example, for finding the best possible pin assignment, we have introduced
the bit-width of the integer in every check statement after assuring that the total
costs of the resulting pin assignment would always be within the scope.

For RQ-3, we consider the \textit{required computation time} as the significantly
influencing factor where researcher and engineers have to cope with when to find
a possible pin configuration during experiments with robotic platforms. Other factors
like memory consumption, experiment preparation time, reusability, or even
model maintenance could have been also considered as influencing the performance.
However, we have agreed on referring to the computation time only in our experiment.

\item \textit{Internal validity.} All experiments were executed on a 1.8GHz Intel
Core i7 with 4GB RAM running Mac OS X 10.8.4. Furthermore, we have used the
same sets of desired configurations  for both RQ-1 and RQ-2. Among them, one
set contains desired configurations of different lengths that are solvable and the
other consists of configurations that are unsolvable.

Concerning both RQ-1 and RQ-2, we outlined a possible solution
how to utilize Prolog and Alloy for model checking. We do not claim having realized
the best solution; yet, our results with respect to the required computation underline
that both approaches are able to handle problem dimensions from real-world examples
in an efficient way to assist researchers and developers.

The results for RQ-3 might be influenced by the chosen execution platform as the
varying computation times Table \ref{tbl:AlloyFeasible-impossibleAssignment}
suggests. However, the standard deviation for these results is rather small and thus,
we consider the negative influence of other running processes on our measurements
to be rather low.

\item \textit{External validity.} As the accompanying search for related work unveiled,
the challenge of solving the pin assignment problem appears to be of relevance for
researchers and developers dealing with robotic platforms, which interact as
cyber-physical systems through sensors and actors with the surroundings. In this regard,
both approaches for RQ-1 and RQ-2 outline useful ways how to address the practical problem of
assigning input sources and output sinks to a hardware/software interface board.
Furthermore, similar combinatorial problems, which can be expressed using either
the graph-based or the tabular representation, can be solved in an analogous manner.

The measurements and results to answer and discuss RQ-3 help researchers and
developers to estimate the computational effort that must be spent to process and
solve problems of a similar size and setup.

\item \textit{Reliability.} Since both outlined solutions for RQ-1 and RQ-2 depend
on the design decisions met by the authors of this article, it is likely that there
might be other designs to realize the model checking approaches in Alloy or
Prolog, respectively. However, according to our results, our design and
implementations are useful enough to be applicable to real-world sized problems.
Since it was not our goal to focus on the utmost optimization for the outlined
design and approaches, future work could be spent in this direction.

With respect to RQ-3, we utilized standardized means to measure the
required computation time. For Prolog, we used its standard profiling
interface \texttt{profile/1} to gather data and for Alloy,
\emph{System.currentTimeMillis()} Java method to calculate the time.
\end{itemize}

\section{Related Work}
\label{sec:RelatedWork}

This article extends our previous work on self-driving miniature vehicles
\cite{BMH13}. Since we are focusing on the software engineering
challenges \cite{BR12a} during the software development for this type of
robotic platforms, this work is aligned with our model-based composable
simulations \cite{BCH+13} where we are trying to find the best suitable
sensor setup for a specific application domain of a robotic platform before
realizing it on the real platform.

The supplier of the STM32F4 Discovery Board provides a tool called MicroXplorer
to assist the developer in verifying the selected pin assignment \cite{STM32-MicroXplorer}.
For that purpose, the user needs to select a desired pin configuration to
let the tool subsequently check wether it is realizable by the microprocessor.
In contrast to that with the verification approaches outlined in this article, we
require the user only to specify the desired set of input sources and output
sinks letting our model checkers finding a feasible, all possible, or the best
pin assignment configuration. Furthermore, our verification approaches are
flexible enough to also enable the merging, concatenation, and comparison
of several existing configurations since both approaches depend only on the
domain model, which can be accessed in a textual way.

Another tool which is freely available is called CoSmart \cite{Coocox} providing
a similar support as the commercial one described before. However, at the time
of writing, our desired hardware/software setup consisting of STM32F4 Discovery
Board and Chibi/OS as real-time operating system is not supported yet.
Moreover, the tool neither assists the user in finding a feasible nor
the best possible pin assignment.

Other work in the domain of model checking using constraint logic programming
was published e.g.~by \cite{GP97} and \cite{GMM90}. They focus on verifying
that a given specification holds certain properties, while our approaches also
aim for optimizing a given combinatorial problem with respect to predefined
costs.

Another approach aiming for utilizing logic programming to find solutions for
a pin assignment configuration problem is reported by the authors of
\cite{JRS12}. However, their work does neither contain a description of a
possible design how to realize this problem using a logical programming
language nor any experimental results.

\section{Conclusion and Outlook}

In this article, we consider the problem of finding \textit{a feasible},
\textit{all possible}, or \textit{the best} pin assignment configuration for
a hardware/software interface board. This task needs to be addressed by
researchers and developers dealing with embedded systems for robotic
platforms to define how a set of sensors like ultra-sonic or infrared
range finders and actors like steering and acceleration motors need to be
connected in the most efficient way.

We have modeled the domain of possible pin configurations for such boards
and analyzed its complexity. On the example of the hardware/software
interface board STM32F4 Discovery Board which we are using on our
self-driving miniature vehicles, we have modeled its pin configuration
possibilities into a graph-based representation. To verify a desired
configuration to be matched with a possible pin assignment, we traversed
the graph and created an equivalent target model for the declarative languages
Prolog and Alloy, respectively. Using our example resulting in 14,689,111
configuration possibilities, we ran an experiment for the aforementioned
three use cases and figured out that Alloy performs up to more than three times
better finding feasible solutions for possible desired configurations and
reporting insolvability of the impossible desired configurations. On the
contrary, Prolog performs up to more than three times better finding all
possible and best solutions for a given desired possible configuration.
Moreover, the Prolog solution is more scalable with the increased configuration
length which is reflected by the lower standard deviations for these use cases.

Using our Eq.~\ref{ConfigurationsM}, it can be seen that the number of possible
configurations increases when either the number of pins or the number of
functions per pin are increased. However, increasing the former let the size of
the problem space grow significantly faster than increasing the latter.
Furthermore, adding more physical pins is also a costly factor; thus,
researchers and engineers will continuously have to deal with the problem of
finding a feasible, all possible, or the best pin assignment configuration for their
specific robotic platform.

Future work needs to be done to analyze this increasing complexity from the
model checking point of view to estimate to which level of complexity instance
models can still be handled properly by the model checking. Furthermore,
semantic constraints like having assigned a pin for data transmission always requires
another pin dealing with data receiving, need to be analyzed how they
constrain the problem space and how they can be considered to optimize
the target models in the particular declarative languages.
The generalizability of the presented approach for finding a pin assignment
configuration in an automated manner needs to be evaluated further with
further popular COTS HW/SW interface boards. The degree of generalizability
would also contribute to determine the effectiveness of the solution for
addressing challenges arising from technical debt.


\addtolength{\textheight}{-12cm}   

\balance

\bibliographystyle{abbrv}
\bibliography{library}



\end{document}